\documentclass[a4paper,twoside]{article}

\usepackage{epsfig}
\usepackage{subcaption}
\usepackage{calc}
\usepackage{tabularx}
\usepackage{amssymb}
\usepackage{amstext}
\usepackage{amsmath}
\usepackage{amsthm}
\usepackage{multicol}
\usepackage{pslatex}
\usepackage{apalike}
\usepackage[bottom]{footmisc}
\usepackage{subcaption}
\usepackage{textcomp}
\usepackage{xcolor}
\usepackage[htt]{hyphenat}
\usepackage[hyphens]{url}
\usepackage{hyperref}
\hypersetup{breaklinks=true}
\usepackage{lstautogobble}
\UseRawInputEncoding
\usepackage{graphicx}
\usepackage{SCITEPRESS}     

\begin{document}

\title{Scalable Infrastructure for Workload Characterization of Cluster Traces}

\author{\authorname{Thomas van Loo\sup{1}, Anshul Jindal\sup{1}\orcidAuthor{0000-0002-7773-5342}, Shajulin Benedict\sup{2}\orcidAuthor{0000-0002-2543-2710}, Mohak Chadha\sup{1}\orcidAuthor{0000-0002-1995-7166}, Michael Gerndt\sup{1}\orcidAuthor{0000-0002-3210-5048}}
\affiliation{\sup{1}Chair of Computer Architecture and Parallel Systems, Technical University Munich, Germany.}
\affiliation{\sup{2}Indian Institute of Information Technology Kottayam, Kerala, India.}
\email{\{thomas.vanloo, anshul.jindal, mohak.chadha\}@tum.de, gerndt@in.tum.de, shajulin@iiitkottayam.ac.in}
}


\abstract{In the recent past, characterizing workloads has been attempted to gain a foothold in the emerging serverless cloud market, especially in the large production cloud clusters of Google, AWS, and so forth. While analyzing and characterizing real workloads from a large production cloud cluster benefits cloud providers, researchers, and daily users, analyzing the workload traces of these clusters has been an arduous task due to the heterogeneous nature of data. This article proposes a scalable infrastructure based on Google's \textit{dataproc} for analyzing the workload traces of cloud environments. We evaluated the functioning of the proposed infrastructure using the workload traces of Google cloud \textbf{cluster-usage-traces-v3}. We perform the workload characterization on this dataset, focusing on the heterogeneity of the workload, the variations in job durations, aspects of resources consumption, and the overall availability of resources provided by the cluster. The findings reported in the paper will be beneficial for cloud infrastructure providers and users while managing the cloud computing resources, especially serverless platforms.}

\onecolumn \maketitle \normalsize \setcounter{footnote}{0} \vfill

\section{\uppercase{Introduction}}
\label{sec:introduction}

In an attempt to invade the minds of cost-conscious users, the cloud providers constantly launch newer execution models or approaches such as ``serverless`` to reduce the cost involved in many applications, especially IoT-enabled applications~\cite{carreira2019cirrus}. There is a shift in the utilization of cloud computing resources such as VMs, monolithic services, microservices, and serverless~\cite{closer20}. This  evolves into varying cloud workloads with complex resource characteristics and requirements for cloud application developers or infrastructure providers. 

Cloud workloads, in general, are classified into two broad classes: i) production jobs that are often latency-sensitive and highly available, and ii) non-production batch jobs that are short-lived and less performance-sensitive jobs~\cite{analysis2011}. These workloads need to be diligently assessed for the better utilization of cloud resources or for enabling a cost-efficient framework. In fact, to achieve cost efficiency, scalability, energy efficiency, and so forth, a few approaches were practiced in the past by cloud infrastructure providers. For instance, approaches such as co-locating suitable serverless functions in the form of establishing a fusion of functions~\cite{costless}, identifying appropriate cloud resources for computations~\cite{espe_closer20}, monitoring the behavior of underneath infrastructures~\cite{arch-specific}, and so forth, have been practiced in the past. 

Identifying appropriate cloud resources, in general, requires a diligent understanding of the existing workloads and their characteristics. However, there are a few challenges for realizing a better performance in the cloud due to the timely characterization of workloads, especially in the evolving cloud markets. A few notable challenges include: 
\begin{enumerate}
\item non-availability of the real-workload traces to predict the resource utilization pattern of clouds; 
\item delayed characterization of workloads; 
\item increasing heterogeneity of resources and workloads -- e.g., FaaS clusters established using Raspberry Pi require minimal computational capabilities when compared to compute clusters established using VMs; and, so forth.
\end{enumerate}

Analyzing the cloud workloads benefits the efficient utilization of cloud resources; besides, it can also lead to effective provisioning of the available heterogeneous compute nodes~\cite{shaju6}. Accordingly, leading cloud providers delivered their workload traces for further workload characterization and analysis -- i.e., Google exposed the traces of workloads carried out at the Borgs' workload for analyzing the cloud resources~\cite{clusterdata:Wilkes2011,clusterdata:Wilkes2020a,clusterdata:Wilkes2020}; Microsoft's Azure constantly updated the traces of its workloads for researchers~\cite{miccrosoft_traces}; Alibaba distributed the CPU utilization of VM workloads of its datacenter~\cite{alibaba_traces}; and, so forth. Although traces are available for further analysis and actions, the existing methods are either not capable of handling larger data traces or inappropriate to handle the heterogeneous nature of workloads. 

In this article, a scalable workload characterization infrastructure based on Google's ``Dataproc`` is proposed~\cite{dataproc:google}. The infrastructure is laid on a spark-based cluster such that the \textit{BigQuery} clients of the architecture analyze the real workload traces of clouds. Experiments were carried out using \textbf{Google cluster-usage traces v3} at our established scalable infrastructure~\cite{clusterdata:Wilkes2020a,clusterdata:Wilkes2020}. In addition, the traces were analyzed from three perspectives: i) analyzing the heterogeneity of the workload traces of the Borg's cluster; ii) characterizing the long or short cloud workloads; evaluating the resource consumption of tasks; and, iii) examining the utilization of cloud resources. 

\textbf{Paper Organization}: Section~\ref{sec:related} discusses the existing research works in the workload characterization domain; Section~\ref{sec:infrastructure} explains the details of the proposed scalable architecture; Section~\ref{sec:results} illustrates the experimental results and associated discussions; and, finally, Section~\ref{sec:conclusion} expresses a few outlooks on the work.

\section{\uppercase{Related Work}} \label{sec:related}
Cloud has remarkably marked its footprints in several research domains, surpassing from IoT to HPC domains~\cite{carreira2019cirrus}. A few research works have been practiced in the past to effectively utilize the cloud resources for varying domains in datacenters and industrial cloud infrastructures. Characterizing cloud workloads has been considered to benefit cloud providers/users due to the cost-effective utilization of resources~\cite{shaju1,shaju5}. Since the first release of the workload trace of a Borg cluster in 2010~\cite{clusterdata:Hellersetein2010}, the cloud community has endeavored to inspect the cloud workload traces in varied fashions to reap in the ultimate insights of the clouds and workload distributions. 

Researchers in the past have analyzed a month-long trace of the single cluster in Google's Borg to study the heterogeneity and dynamicity of the workloads~\cite{10.1145/2391229.2391236,6779441,8450304}. Authors of~\cite{10.1145/2391229.2391236} studied the challenges due to the inclusion of heterogeneous workloads and inefficient resource scheduling aspects of cloud resources. A few authors attempted to apply machine learning algorithms to assess the workloads of traces. For instance, the authors of~\cite{analysis2011} and~\cite{9209730} have utilized the data traces to predict the workload requirements when executed on the cloud environments. The authors clustered the cloud workloads of similar patterns after developing a machine learning model. 

The dataset, we analyze in this work~\cite{clusterdata:Wilkes2020a,clusterdata:Wilkes2020}, was released in April 2020. The main addition to the third google cluster trace compared to the second was the tracing of 8 separate clusters spread out through North America, Europe, and Asia instead of the measurements of a single cluster as provided in the second trace. Furthermore, three additions have been made to the properties recorded: CPU usage information histograms are provided every 5 minutes instead of the previous single point sample. Alloc sets were equally introduced to the third trace, which was not present in the 2011 dataset. Lastly, job-parent information for master/worker relationships, such as MapReduce, has been added. It is also worth noting that at approximately 2.4TiB of data, this trace is far more significant than the previous recordings. Due to its size, the third trace has only been made publicly available in Google's cloud data warehouse, BigQuery~\cite{bigquery}.  

In previously existing approaches, the analysis has been executed on a single machine or has not considered the recent cloud data traces. Thus, this work proposed the scalable workload characterization environment to evaluate the cloud workloads. 

\section{\uppercase{Infrastructure}} \label{sec:infrastructure}
\begin{figure}[t]
  \centering
  \includegraphics[width=0.95\linewidth]{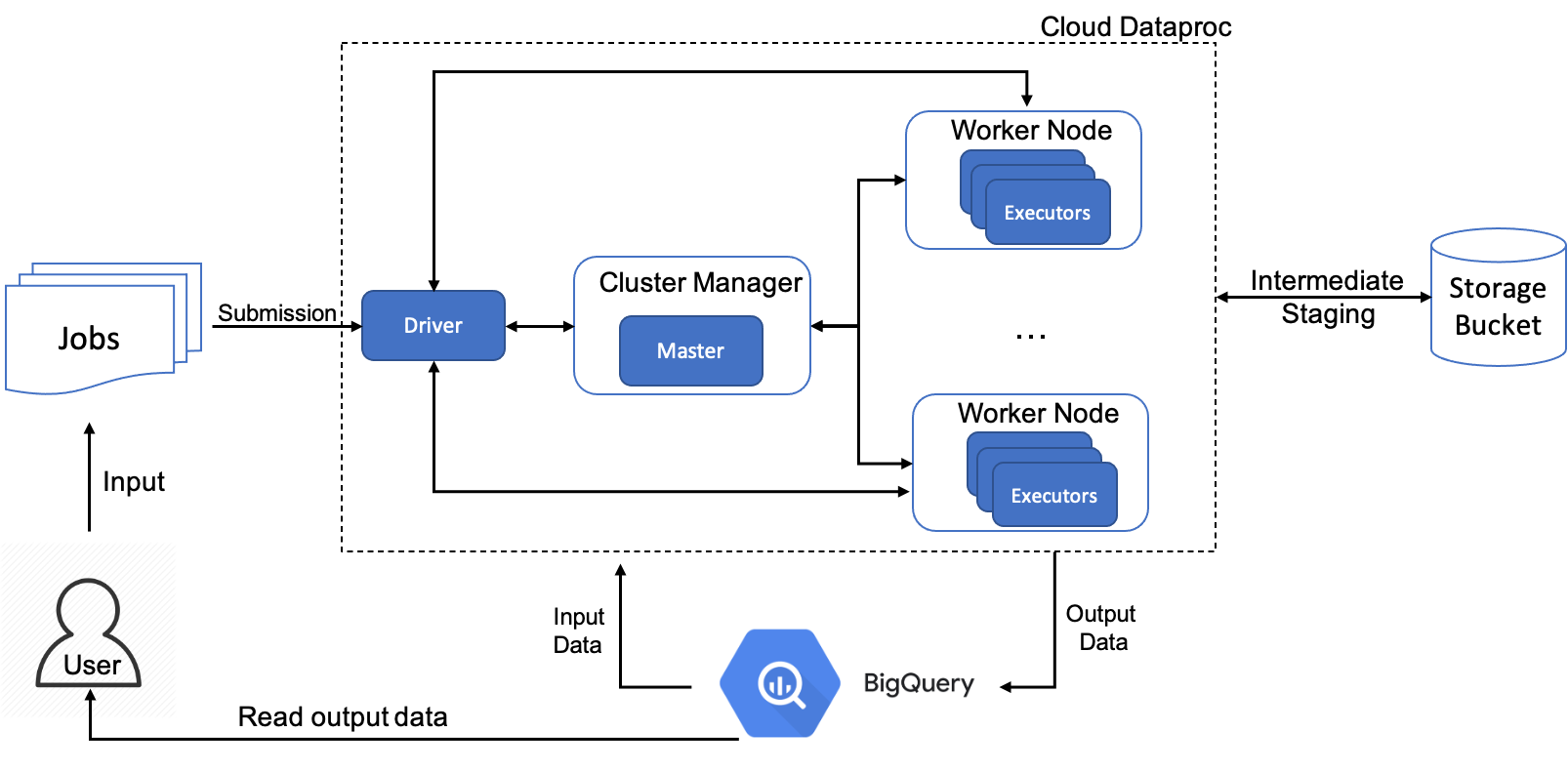}
  \caption{Scalable Infrastructure based on Google Cloud Dataproc for Workload Characterization} \label{fig:scalable_archi}
\end{figure}

To handle large volume of cloud workload traces and analyze the characteristics of them, we proposed a scalable environment consisting of Dataproc\footnote{https://cloud.google.com/dataproc} from the Google cloud. The proposed scalable infrastructure paved way for efficient data analytics processing much faster than the other approaches. A high level overview of the proposed infrastructure and workflow is shown in Figure~\ref{fig:scalable_archi}. The important entities that are available in the proposed scalable architecture and their functionalities are described in the following subsections.

\subsection{YARN-Cluster}
To cope with the evolving range of cloud workload traces, we included Yet Another Resource Negotiator (YARN) cluster in the architecture using Google's ``Dataproc`` component. YARN, in general, can handle big data for analytics purposes. It is lightweight due to the metadata model of establishing clusters. Firstly, a YARN (Yet Another Resource Negotiator)  based spark cluster is created using the Dataproc. This cluster consists of one master node and ``n`` worker nodes, where n is equal or greater than 2. The master node manages the cluster, creating multiple executors within the worker nodes. These executors are responsible for running the tasks in parallel. The type of the virtual machine used for the master and worker nodes can be specified along with the necessary libraries required for the job while creating a cluster. Once the cluster is created, the libraries are installed automatically and are configured to be linked with external \textit{Storage Units} which the executors use.

\subsection{Storage Units}
The proposed scalable architecture includes external storage units -- e.g., google cloud storage buckets -- for processing the workload traces. These storage units are responsible for performing intermediate data staging while executing the tasks. In most cases, these are required for executing the analysis tasks. 

\subsection{Job Submissions}
Analyzing cloud workload traces is carried out by submitting the jobs to the scalable infrastructures via the job submission portal. The Job Submission portal is designed using PySpark or Spark-based job description system connected with the BigQuery component. The primary purpose of including the BigQuery component in the system is to enable the executors to access data directly from the BigQuery. In doing so, the users could view or utilize the environment to further process traces. Additionally, the Job Submission portal is designed to establish an autoscaling feature of worker nodes to make the analysis of workloads scalable. 

\subsection{Dataset Access}
A more appropriate dataset with suitable cloud workload information is crucial for characterizing the cloud jobs. The proposed scalable architecture applied the most recent Google Cluster-Usage traces v3 for the characterization of the clouds. Accessing cloud traces in the proposed scalable architecture is handled through the BigQuery-based serverless platform. In general, BigQuery is a fully-managed, serverless data warehouse that enables scalable, cost-effective and fast analysis over petabytes of data. It is a serverless  Software-as-a-Service (SaaS) that uses standard SQL by default as an interface~\cite{bigquery}. It also has built-in machine learning capabilities for analyzing the dataset capabilities.  
There are various available methods to access data within BigQuery, including a cloud console, a command line tool and a REST API. In this work, we preferred BigQuery API client library based on Python due to the inherent capabilities of processing machine learning features.

\section{\uppercase{Results}}\label{sec:results}
In this section, we first discuss the workload characterization from four different aspects: i) Heterogeneity of collections and instances (\S\ref{sec:heterogeneity_col_inst}), ii) Jobs' duration characterization (\S\ref{sec:jobs_dur_char}), iii) Tasks' resources usage (\S\ref{sec:tasks_res_usage}), and iv) Overall cluster usage (\S\ref{sec:overall_clust_usage}). 

\subsection{Heterogeneity of properties}
\label{sec:heterogeneity_col_inst}
This part of the analysis is focused on the heterogeneity of the properties within the \textbf{CollectionEvents} and \textbf{InstanceEvents} tables, as well as examining and visualizing the quality of the dataset. We will go property by property, commencing with those that are common within both tables (\S\ref{sec:common_properties}), then the ones specific to CollectionEvents (\S\ref{sec:collect_events_specific}) and lastly specific to InstanceEvents (\S\ref{sec:instance_events_specific}).

\subsubsection{Common Properties}
\label{sec:common_properties}
\begin{figure*}[t]
\centering
\begin{subfigure}[b]{0.49\linewidth}
  \centering
  \includegraphics[width=0.9\linewidth]{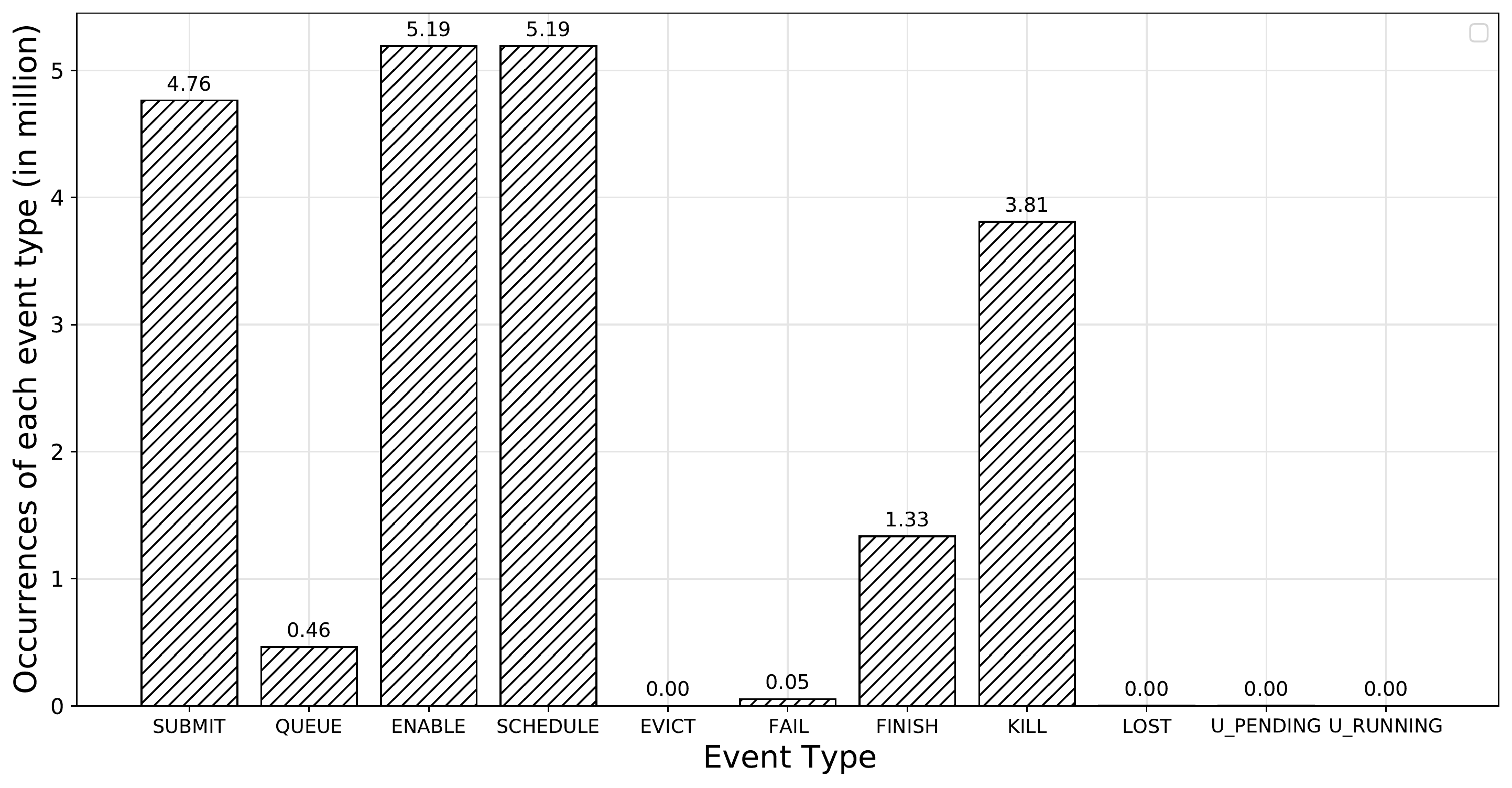}
  \caption{\textbf{CollectionEvents} table} \label{fig:collec_types}
\end{subfigure}%
\hfill
\begin{subfigure}[b]{0.49\linewidth}
  \centering
  \includegraphics[width=0.9\linewidth]{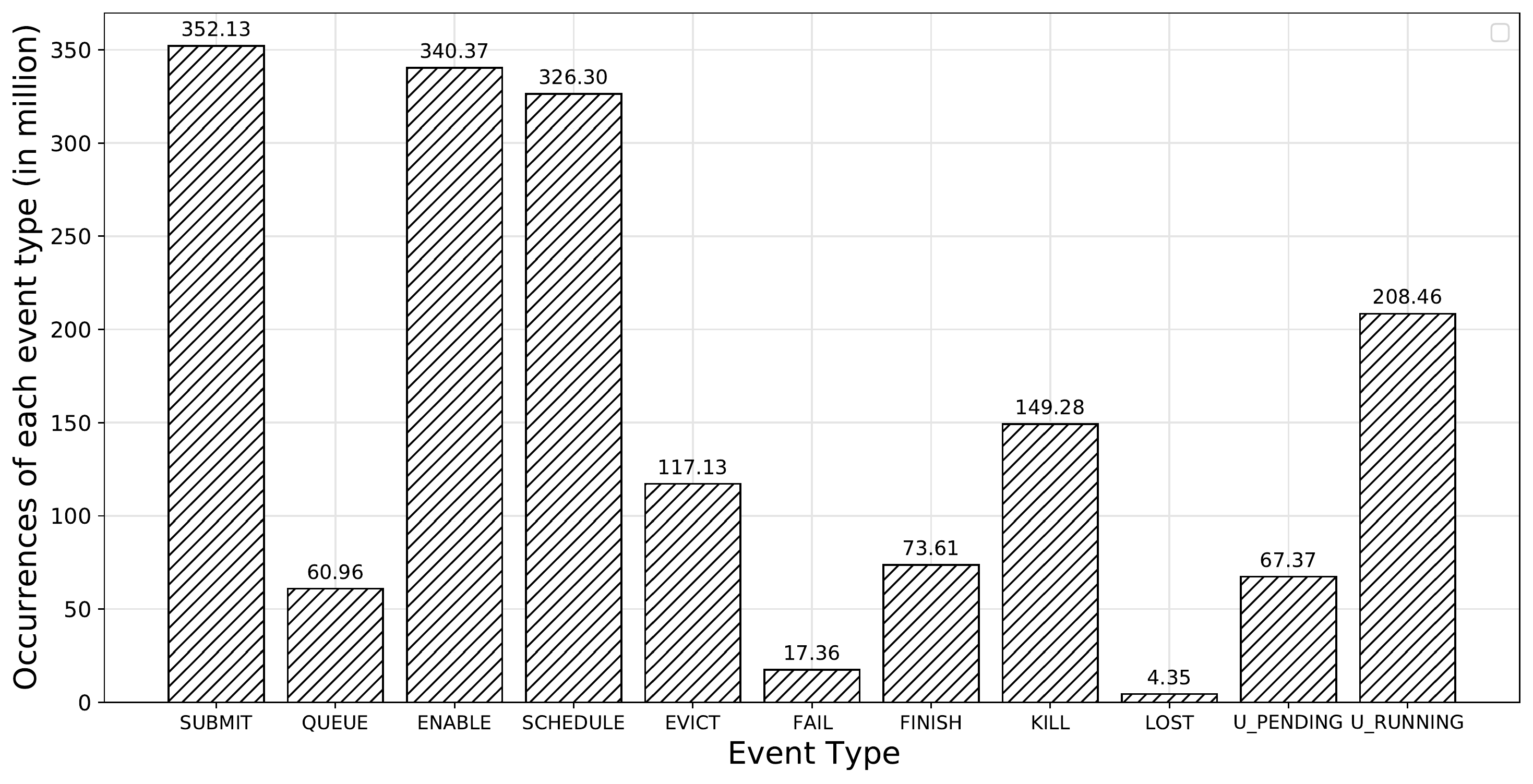}
  \caption{\textbf{InstanceEvents} table} \label{fig:inst_types}
\end{subfigure}
\caption{
Number of occurrences of different event types within two major types of tables.}
\label{fig:inst_types_coll}
\end{figure*}

We begin by examining the total amount of collections and instances submitted throughout the trace period to gain a broader overview of the spectrum of the tables. There are a total of approximately \texttt{20.1} million rows in the \textbf{CollectionEvents} table, which represent events that occur on the roughly \texttt{5.2} million unique \textit{collection\_id}'s present. Over \texttt{99}\% of these are jobs, with roughly \texttt{36} thousand alloc sets. The \textbf{InstanceEvents} table comprises roughly \texttt{1.7} billion rows, which represent the number of instances spread out over the \texttt{5.2} million collections. Approximately \texttt{87}\% of the instances are tasks and roughly \texttt{13}\% alloc instances. We further investigate the distribution of the occurrences of the different event types throughout the dataset. The count for each type in the table is displayed in Figure~\ref{fig:collec_types}. 

\begin{figure*}[t]
\centering
\begin{subfigure}{1\columnwidth}
    \centering
        \includegraphics[width=\columnwidth]{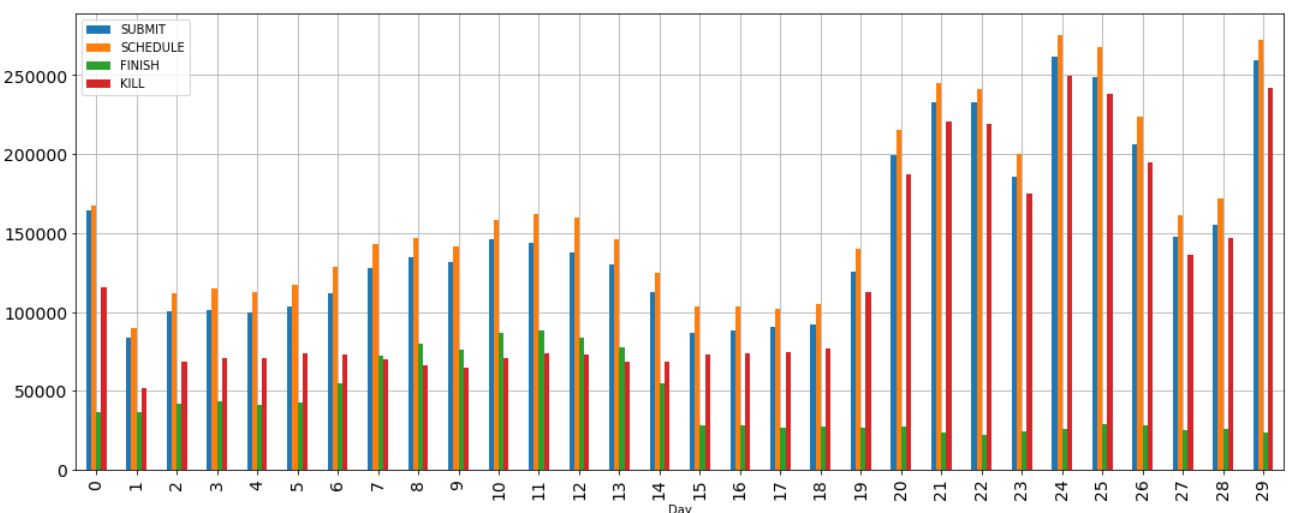}
        \caption{Occurrences of \textit{SUBMIT}, \textit{SCHEDULE}, \textit{FINISH} and \textit{KILL} events in \textbf{CollectionEvents} per day.} \label{fig:collec_typespd}
\end{subfigure}\hfill
\begin{subfigure}{1\columnwidth}
    \centering
        \includegraphics[width=\columnwidth]{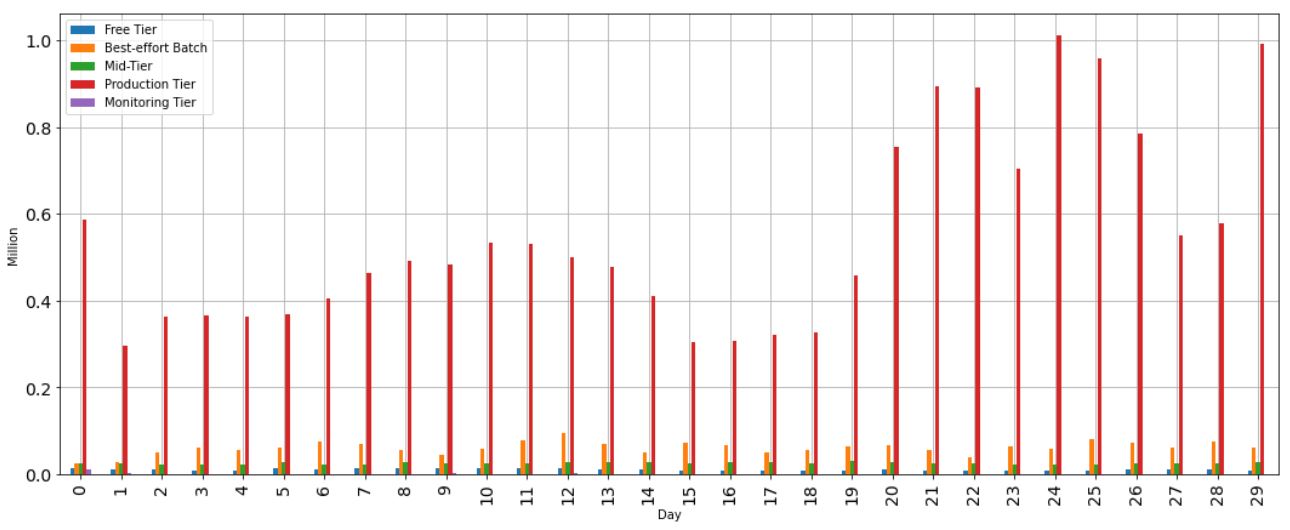}
  \caption{Counts of events associated with the 5 priority tiers per day.} \label{fig:priorities_pd}
\end{subfigure}\hfill
\caption{
Occurrences of events during the course of trace collection time.}
\label{fig:high_user_workload}
\end{figure*}

From Figure~\ref{fig:collec_types} we can see that, the most common events are \textit{SUBMIT}, \textit{ENABLE}, \textit{SCHEDULE}, \textit{FINISH} and \textit{KILL} which were to be expected. As an additional step, we count the number of collection events that are associated with these events per day (excluding \textit{ENABLE} as this is comparable to \textit{SCHEDULE} in this context). Figure~\ref{fig:collec_typespd} shows the result. We find from the result that: (1) the number of scheduled collections shows a certain periodicity to an extent as every \texttt{14} days, the count falls for several days. (2) Collections are frequently killed, far more often than they finish normally. (3) During the last nine days of the trace, there was a surge of activity, with almost double the amount of submitted and scheduled collections.

Collections and instances are equipped with a \textit{priority} property: a small integer, with higher values indicating higher priorities. Those with more significant priorities are given preference for resources over those with smaller priorities. The values can be sub-categorized into 5 tiers and the distribution of jobs within these 5 tiers is shown in Table~\ref{prio_table}, with the production tier as the clear majority.

\begin{table}[t]
\centering
\caption{Occurrences of collection scheduling classes.}
\begin{tabularx}{\columnwidth}{|X|l|}
        \hline
        \textbf{Tier} & \textbf{Percentage of all jobs} \\
        \hline
        Free & 1.6\% \\
        \hline
        Best-effort Batch & 9.1\% \\
        \hline
        Mid & 3.8\% \\
        \hline
        Production &  85.2\% \\
        \hline
        Monitoring & 0.3\%\\
        \hline
    \end{tabularx}
    \label{prio_table}
\end{table}

As with the event types, we display the number of events associated with each priority tier for each day. We can see in Figure~\ref{fig:priorities_pd}, the production and best-effort batch tier jobs follow the same trends as the event types throughout the trace, contrary to the free, mid, and monitoring tier jobs, which show no discernible patterns.

The last property common to both tables is \textit{alloc\_collection\_id}, the id of the alloc set that hosts the job, or empty if it is a top-level collection. Upon investigating this property, we found roughly \texttt{97.7}\% of jobs to be top-level collections and a mere \texttt{0.3}\% to be hosted by alloc sets.

\subsubsection{CollectionEvents -- Specific Properties}
\label{sec:collect_events_specific}
The first property we examine that is unique to \textbf{CollectionEvents} is \textit{parent\_collection\_id}, the ID of the collection's parent or an empty value if it has none. We found that approximately \texttt{64}\% of all collections had parent ID, indicating that most collections are related to others.

The next distribution we investigate is that of the \textit{max\_per\_machine}, the maximum number of instances from a collection that may run on the same machine. About \texttt{99}\% of all collections do not have a constraint in this regard. For the remaining \texttt{1}\% with a \textit{max\_per\_machine} value, there are four unique values that exist in the table. These values and their counts are displayed in Table~\ref{mpm_table}. 

\begin{table}[t]
\centering
\caption{Occurrences of \textit{max\_per\_machine} values.}
\begin{tabularx}{\columnwidth}{|X|l|}
        \hline
        \textbf{Max per Machine} & \textbf{Count} \\
        \hline
        1 & 35150 \\
        \hline
        2 & 289 \\
        \hline
        10 &  2 \\
        \hline
        25 & 36\\
        \hline
    \end{tabularx}
    
    \label{mpm_table}
\end{table}

From these occurrences, we can see that should a collection come with a constraint for instances running on a single machine, around \texttt{99}\% of the time, it will be limited to \texttt{1} instance per machine. Similar to the previous property, \textit{max\_per\_switch} displayed similar tendencies, with over \texttt{99}\% of collections not having any constraints. The few that had a value for this field amounted to \texttt{21} unique values ranging from \texttt{1} to \texttt{104}. \texttt{99}\% of these collections were limited to one instance per switch.

Users have the option of enabling \textit{vertical\_scaling} when submitting a job, allowing the system to decide how many resources are required autonomously. The dataset displays this information via four unique values in this field. Table~\ref{vsd_table} shows the result of the examination of the distribution of the values for \textit{vertical\_scaling}. We see that most users have enabled vertical scaling, most of which are bound by user-specific constraints. We assume the constraints are widely used to ensure cost-effectiveness when consuming cloud resources.

\begin{table}[t]
\centering
\caption{Distribution of values for \textit{vetical\_scaling}}
\begin{tabularx}{\columnwidth}{|X|l|}
        \hline
        \textbf{\textit{vertical\_scaling} value} & \textbf{Percentage of Jobs} \\
        \hline
        Setting Unknown & 0.0\%\\
        \hline
        Off & 6.8\% \\
        \hline
        User-Constrained & 66.6\%  \\
        \hline
        Fully Automated & 26.6\% \\
        \hline
    \end{tabularx}
    \label{vsd_table}
\end{table}

\subsubsection{InstanceEvents -- Specific Properties}
\label{sec:instance_events_specific}
We begin the \textbf{InstanceEvents} specific properties by analyzing the \textit{instance\_index} column, which indicates the position of an instance within its collection. Using this information, we can determine variations in the number of tasks within jobs. The maximum value in the table for this field is \texttt{97,088}. In Table~\ref{jobtasks}, we display the job size distribution in terms of tasks per job, separated into \texttt{6} bins. We can see that jobs primarily have several tasks under ten and rarely over \texttt{1000}. 

\begin{table}[t]
\centering
\caption{Distribution of job sizes in terms of tasks per job.}
\begin{tabularx}{\columnwidth}{|X|l|}
        \hline
        \textbf{Number of Tasks} & \textbf{Number of Jobs} \\
        \hline
        1 & 4,067,109 \\
        \hline
        2 - 10 & 906,736 \\
        \hline
        11 - 100 & 149,516  \\
        \hline
        101 - 1000 & 72,984 \\
        \hline
        1001 - 2000 & 7,715 \\
        \hline
        $>$ 2000 & 9,606 \\
        \hline
    \end{tabularx}
    \label{jobtasks}
\end{table}

Furthermore, we examine the \textit{machine\_id} property that provides the ID of the machine an instance was scheduled on. We found that roughly \texttt{51.9}\% of all instances had been scheduled on a machine, while the remaining \texttt{48.1}\% are yet to be scheduled.









\subsection{Job Duration Characterization}
\label{sec:jobs_dur_char}
In this section, we first discuss the overall characterization of short and long job's durations and their success rates (\S\ref{sec:job_dur_success}).
Then we analyze the jobs' durations and success rates by priority tier (\S\ref{sec:job_dur_success_prio}). Lastly, we analyze the variations of time spent in different states of a jobs' lifecycle (\S\ref{sec:job_state_dur}).

\subsubsection{Overall jobs' durations and success rates}
\label{sec:job_dur_success}
We measure job durations in seconds, the longest possible duration in the dataset being around \texttt{2,600,000s} seconds (\texttt{30} Days). Further, we found that roughly \texttt{85.3}\% of all jobs had a duration between \texttt{0s} and \texttt{1000s}, only around \texttt{0.4}\% run longer than one day, and a total of \texttt{48} jobs runs during the entire trace period. Of the jobs that ran under \texttt{1000s}, around \texttt{60}\% had a run time under \texttt{100s}, which is around half of the jobs. 

\subsubsection{Jobs' durations \& success rates by priority}
\label{sec:job_dur_success_prio}
In this section, we analyze the jobs' durations and their success rates by priority tiers. The success rates of jobs in all the tiers is shown in Table~\ref{sr_free}. 
\begin{table}[t]
\centering
    \caption{Success rates of jobs in all the tiers.}
    \resizebox{\columnwidth}{!}{
    \begin{tabular}{|l|l|l|l|l|l|}
        \hline
        \textbf{Final State} & \textbf{Free }& \textbf{BEB} & \textbf{Mid}& \textbf{Prod.}& \textbf{Monit.}\\
        \hline
        KILL & 49.2\% & 58.9\% & 49.5\% & 76.5\% & 9.2\%\\
        \hline
        FINISH & 47.4\% & 36.9\% & 48.9\% & 23.1\% & 4.4\% \\
        \hline
        FAIL & 3.4\%  & 4.2\% & 1.6\%  & 0.4\%  & 81.3\%  \\
        \hline
        SCHEDULE & & & & & 5.1\% \\
        \hline
    \end{tabular}
    }\label{sr_free}
\end{table}

\textbf{free tier}: The overall jobs' durations distribution is analogous to the overall distribution, with the run times under, \texttt{1000s} and the majority of those jobs run under \texttt{100}. From Table~\ref{sr_free}, we see that the kill to finish ratio is more balanced here, being almost equally distributed.

\textbf{best-effort batch tier}: They showed slightly elevated levels of jobs that ran longer than \texttt{1000s} in comparison to the overall distribution, yet had no jobs running longer than a day. This is to be expected, as \textit{best-effort batch tier}  was conceived for jobs handled by the batch scheduler. The mean run time for  \textit{best-effort batch tier}  is roughly \texttt{7227s}. From Table~\ref{sr_free}, we observe that the success rates for \textit{best-effort batch tier} are less evenly spread in comparison to the \textit{free tier}, with a higher KILL rate. This suggests that longer running jobs could be killed more frequently than shorter running ones.

\textbf{mid tier}: They also displayed a similar run time distribution as the overall one, except having a slightly more elevated count of jobs that lasted between \texttt{1000s} and \texttt{2000s}. The mean job duration within this category is approximately \texttt{6653s}. From Table~\ref{sr_free}, the success rate is evenly distributed among FINISH and KILL, a further indication that longer jobs tend to be killed more often than shorter ones.

\textbf{production tier}: They displayed a very similar pattern to the overall free and mid tier distributions, consisting heavily of jobs with runtimes under \texttt{1000s} and having a mean of about \texttt{1885s}. This indicates that production jobs, which have the highest priority for everyday use, mainly consist of short-running jobs that ran under \texttt{100s}. From Table~\ref{sr_free}, contradictory to our previous assumption, production jobs show an apparent tendency to be killed more frequently than not, despite consisting almost exclusively of very short-running jobs.

\textbf{monitoring tier}: As expected, with a mean of \texttt{42,970s}, the runtimes of jobs in the monitoring tier are, on average, the longest of all tiers. The large majority of them ran between \texttt{3000s} and \texttt{4000s}, but this group was also the only one with visible counts for runtimes over \texttt{50,000s} seconds. The success rate for this tier, shown in table~\ref{sr_free}, is characterized by a resounding majority of jobs failing, a clear indication that long-running jobs have a much higher tendency to fail than shorter ones. Some jobs last state were also recorded as SCHEDULE. We expect these jobs ran during the entire trace period without finishing.

\begin{figure}[t]
  \centering
  \includegraphics[width=0.7\linewidth]{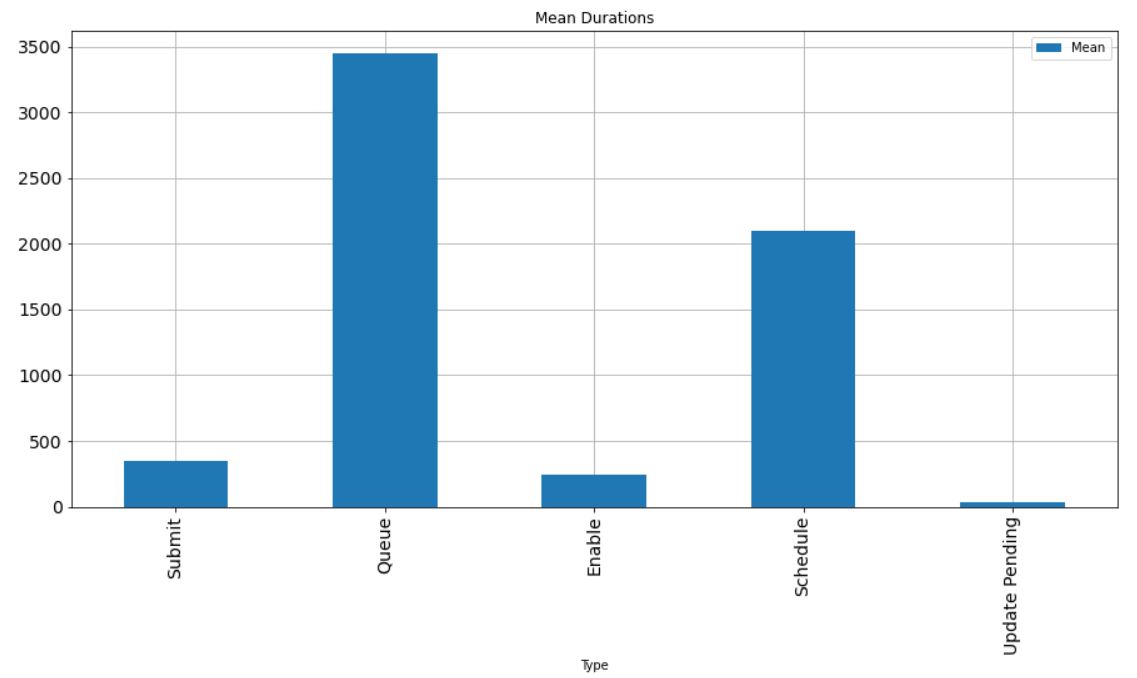}
  \caption{Mean durations of jobs spent in each state.} \label{fig:mean}
\end{figure}
\subsubsection{Job State Durations}
\label{sec:job_state_dur}

\begin{figure*}[t]
\centering
\begin{subfigure}[b]{0.49\linewidth}
  \centering
  \includegraphics[width=0.9\linewidth]{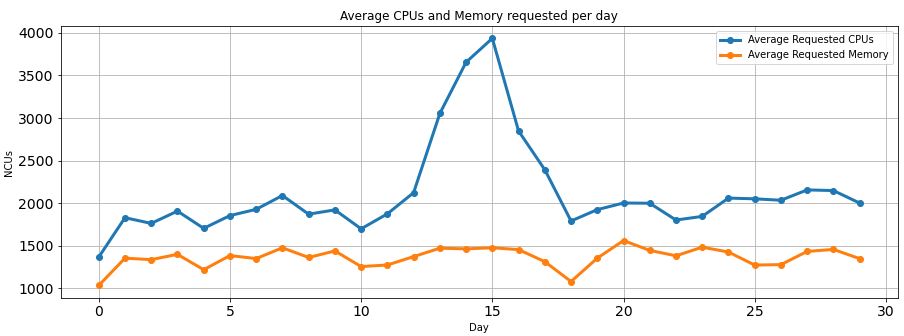}
  \caption{Requested} \label{fig:avgreqs}
\end{subfigure}%
\hfill
\begin{subfigure}[b]{0.49\linewidth}
  \centering
  \includegraphics[width=0.9\linewidth]{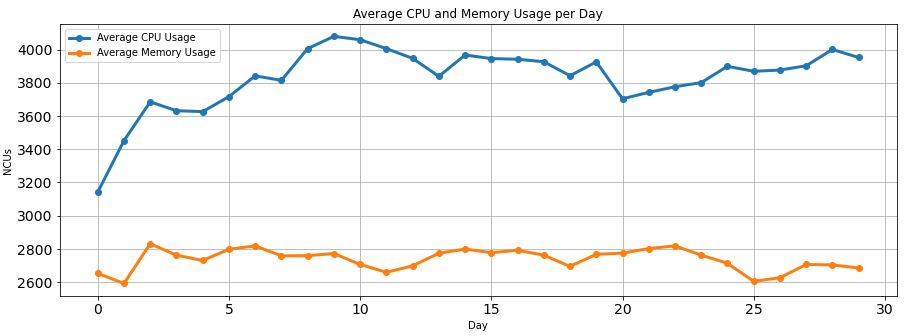}
  \caption{Consumed} \label{fig:avgusage}
\end{subfigure}
\caption{Average CPU and memory requested and consumed per day.}
\label{fig:task_resources_request_dist}
\end{figure*} 

This section analyzes the time spent by jobs in different states, specifically SUBMIT, QUEUE, ENABLE, SCHEDULE, UPDATE RUNNING and UPDATE PENDING. The other four states mark the end of a job's life cycle and do not have a duration. We do this by examining these six states' mean. It is to be noted that these events are not actually states, but events that trigger transitions between job states; however, by determining the elapsed time between these transitions, we can calculate how much time was spent in the respective state. For this reason, we refer to these events as states in this section.

Figure~\ref{fig:mean} shows the mean durations of jobs spent in each state. With around \texttt{424,477s}, UPDATE\_RUNNING is the state with the most extended value by a large margin. The mean values for the other states are barely visible in the comparison graph. We, therefore, display the graph without the value for UPDATE RUNNING in Figure~\ref{fig:mean}, allowing a more apt comparison. This state also had the lowest occurrence, having only a count of \texttt{20} jobs that ran an update during the trace. Though updates rarely happen, we can conclude that they tend to be much more time costly.

From Figure~\ref{fig:mean}, we further see that the second-highest value is for the QUEUE state. Upon further investigation, we found that the majority of jobs spent under \texttt{100s} in this state, yet numerous outliers are ranging from over \texttt{1000s} till a maximum of around \texttt{1,200,000s}, which explains the higher mean value. We suspect these are jobs in the lower priority tiers that have to remain in queue until higher priority jobs yield compute resources to run on. 

From the three remaining states, we conclude that: (1) The time lapses between a job being SUBMITTED and either ENABLED or QUEUED is relatively low, which shows the time between submission and eligibility to be scheduled is kept at a minimum. (2) The low mean of ENABLE state signifies that once a job becomes eligible for scheduling, it does not take long for the scheduler to place it compared to the rest of the job's life cycle. (3) When a job needs to be updated, the pending time before the update is performed is meager, unlike the average update time.

\subsection{Task Resource Usage}
\label{sec:tasks_res_usage}
In this section, we first analyze the amount of resources that are requested by tasks, which is saved in the \textit{resource\_request} column of the \textbf{InstanceEvents} table as a structure (\S\ref{sec:tasks_resource_requests}). It represents the maximum CPU or memory a task is permitted to use. Second, we analyze the amount of resources consumed by tasks, which is recorded in the \textbf{InstanceUsage} table (\S\ref{sec:tasks_resource_usage}). These two aspects can then be compared to view the general resource requirements for a cluster to handle this type of workload and see if tasks resource limits are adhered to or exceeded. 

\subsubsection{Resource Requests}
\label{sec:tasks_resource_requests}

We calculate the average resource requests and later usage per day based on this tracing system to accurately compare the resource requests with the average resource usage. We separate the entire trace duration timeline into windows of \texttt{5} minutes, giving us \texttt{288} windows per day. We then determine the sum of all resource requests per window and calculate the average value of the \texttt{288} sums for each day, representing the average request values of that particular day. The result is plotted in Figure~\ref{fig:avgreqs}. The immediate observation is that the average CPU request spikes and reaches its maximum value on day \texttt{15}. This is interesting, as day \texttt{15} is also the day with the minimum job activity in the whole trace period. This could indicate that the job count for that day is so low that the jobs submitted had tasks that were expected to be CPU intensive, causing the allowed number of jobs to be submitted to drop.

\subsubsection{Resource Usage}
 \label{sec:tasks_resource_usage}
We first determine the average CPU and memory consumption by tasks per day to perform the comparison. The way these values are calculated is analogous to that of the average resource request, the exception being that instead of summing the total amount of requests per trace window, we use the sum of all the average task consumptions recorded in each window. The average value for each day is then given as the average of the 288 usage sums on that day. These average values are shown in Figure~\ref{fig:avgusage} for both CPU and memory values. Contrary to the resource requests, the average CPU usage does not reach its maximum value on day \texttt{15}. We expect this abnormality is due to the exception values described previously, where a request value is set to \texttt{0}, which implies it is not set a limit to the resources it may use on some machines. These values would lead to a higher consumption on some days instead of the average requests, which would be lowered.



\subsection{Overall Cluster Usage}
\label{sec:overall_clust_usage}
\begin{figure}[t]
  \centering
  \includegraphics[width=0.7\linewidth]{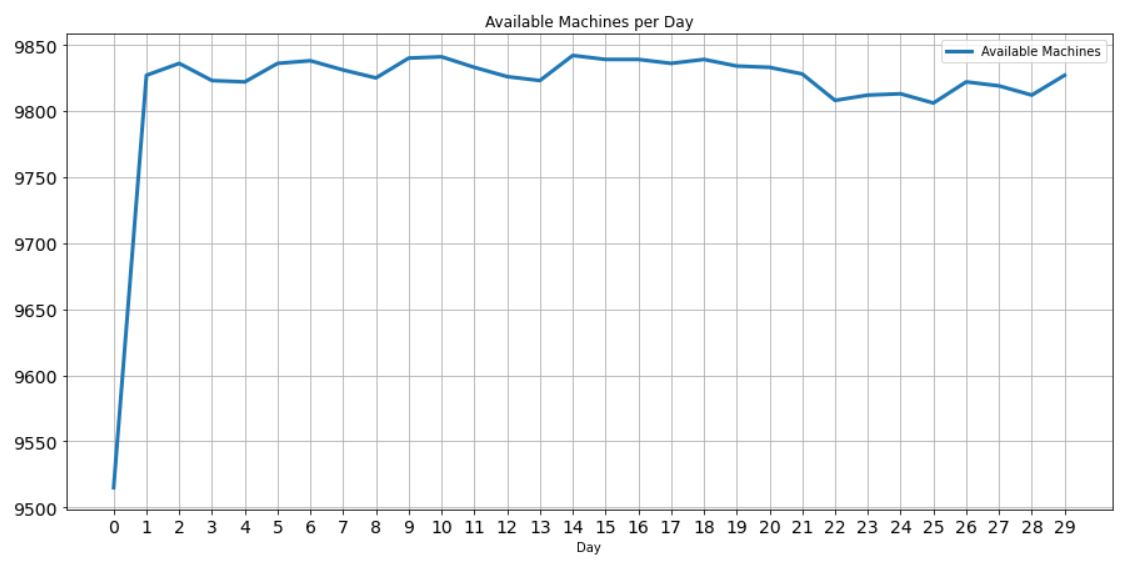}
  \caption{Available machines per day.} \label{fig:machavail}
\end{figure}

\begin{figure*}[t]
\centering
\begin{subfigure}[b]{0.485\linewidth}
  \centering
  \includegraphics[width=1\linewidth]{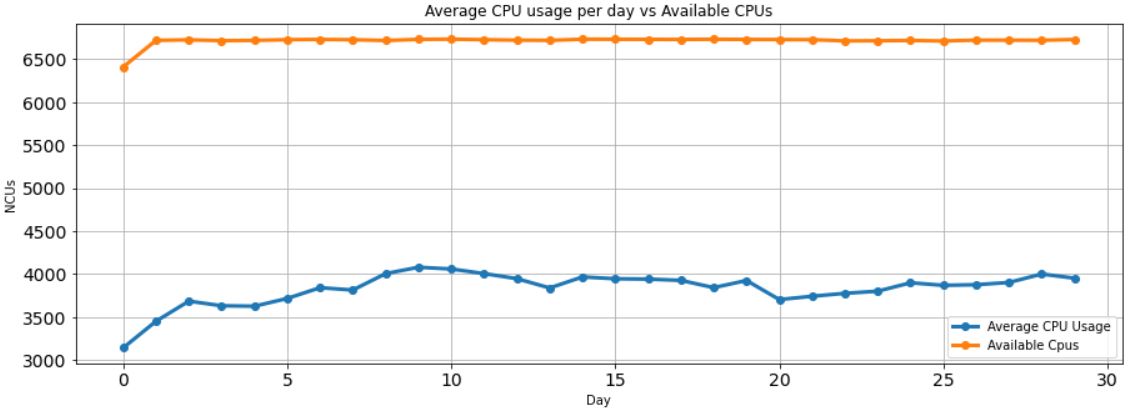}
  \caption{Average CPU usage vs Availability per day.} \label{fig:cpu_uvsa}
\end{subfigure}%
\begin{subfigure}[b]{0.485\linewidth}
  \centering
  \includegraphics[width=1\linewidth]{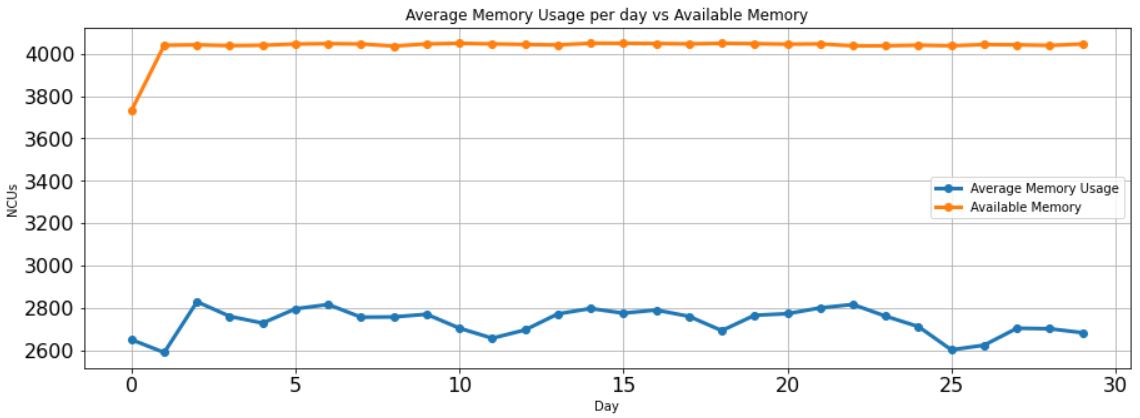}
  \caption{Average Memory usage vs Availability per day.} \label{fig:mem_uvsa}
\end{subfigure}
\caption{
Average usage of resources vs their availability per day.}
\label{fig:usage_available}
\end{figure*}

We perform the overall cluster usage analysis by examining the \textbf{MachineEvents} table to see how many machines are available to the cluster, the rate at which they are added and removed, how many resources are readily available to jobs and how much of those resources are used.  We assume the number will remain relatively consistent throughout the trace. The result can be seen in Figure~\ref{fig:machavail}, confirming our prediction. We can see that the cell we are analyzing has between \texttt{9800} and \texttt{9850} machines available every day throughout the trace.


We also calculate the normalized sums of the available CPU cores and RAM sizes that these machines provide, comparing with the resource usage of tasks, showing us how much of these are actively used. The comparisons are shown in Figure~\ref{fig:cpu_uvsa} and Figure~\ref{fig:mem_uvsa}. The units in these graphs are normalized compute units~\cite{clusterdata:Wilkes2020a}. The first observation is that the available CPU and memory resources stay consistent, despite the frequent additions and removals of machines per day. Furthermore, as we can see, the average resource consumption by tasks never reaches the full capacity of the cell. The maximum average CPU usage occurs on day nine and uses roughly \texttt{60}\% of the total capacity. For memory, the maximum average usage takes place on day two of the trace and consumes around \texttt{70}\% of the total capacity. Therefore, in general, this cell has around at least 40\% of its compute power lying idle daily and at least around \texttt{30}\% of its memory capacity unused.

\begin{figure}[t]
  \centering
  \includegraphics[width=0.7\linewidth]{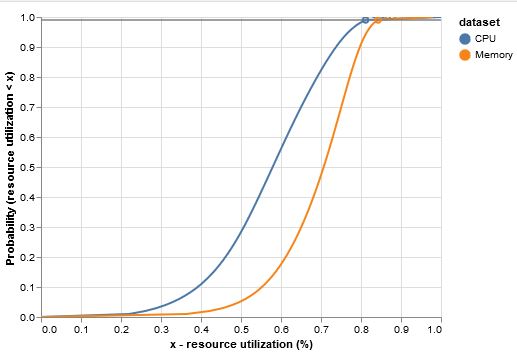}
  \caption{Cumulative graph displaying the probability of the cell's resource utilization being at most x.} \label{fig:cumulative}
\end{figure}

As an additional evaluation, we present a cumulative graph of the probability of overall consumption of CPU and memory resources in Figure~\ref{fig:cumulative}. The x-axis in this graph displays the percentage of the total resources utilized. The y-axis shows the probability of the utilization being at most x at any given point in time during the trace. As we can observe, the probabilities rise drastically as of around \texttt{40}\% total resource utilization, with the maximum probability being reached at around \texttt{80}\%. We gather from this, that at no point in time during the trace was there more than \texttt{80}\% of the total capacity of the cell being used.

\subsection{Infrastructure scalability evaluation}\label{sec:scalability_eval}

\begin{figure}[t]
  \centering
  \includegraphics[width=0.8\linewidth]{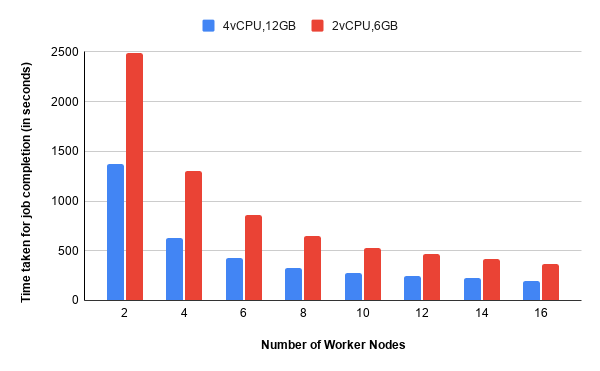}
  \caption{Dataproc-based infrastructure scalability evaluation considering two types of virtual machines cluster having different compute resources. } \label{fig:scalability_eval}
\end{figure}

We have evaluated the scalability of the Dataproc-based Infrastructure by determining the time taken by the \textit{resource-usage} calculation job on two different types of machines cluster: \texttt{machine1} with 4 vCPU and 12GB memory, and \texttt{machine2} with 2 vCPU and 6GB memory.
Figure~\ref{fig:scalability_eval} shows the evaluation results for the two types of machines with a different number of worker nodes. It can be observed that the time required for characterizing the workload traces steadily decreases with the number of worker nodes for both types of machines. Additionally, the values become almost constant as the worker nodes reach greater than \texttt{12}.
Notably, \texttt{machine2} with 4, 8, 12 and 16 worker nodes have almost the same completion time as the \texttt{machine1} at 2, 4, 6 and 8 worker nodes respectively. For instance, around 280 seconds is observed when worker nodes are above 8. This is attributed to the fact that both clusters have approximately the same number of cores and memory, leading to the same performance.  

\section{\uppercase{Conclusion}} \label{sec:conclusion}
Hosting workloads considering the heterogeneous nature of cloud characteristics has been a pivotal point of research for several cloud researchers and infrastructure providers in the recent past. A diligent characterization of workloads eases infrastructure providers and cloud developers. However, an efficient characterization approach of traces for the modern cloud execution models such as serverless functions is not undertaken in the past. In this work, we studied the \textbf{Google cluster-traces v3} dataset, the latest of the Google cluster traces, by analyzing its properties and performing a workload characterization of the traces using our proposed scalable infrastructure based on Google Cloud Dataproc. We perform the workload characterization on this dataset, focusing on the heterogeneity of the workload, the variations in job durations, aspects of resources consumption, and the overall availability of resources provided by the cluster. Furthermore, we also show the scalability analysis of the proposed infrastructure. The findings reported in the paper will be beneficial for cloud infrastructure providers and users while managing the cloud computing resources, especially serverless platforms. 

In the future, we will further analyze missing insights of the workload traces using our scalable infrastructure to study properties such as page cache memory, CPI, and CPU usage percentiles to provide further insights regarding the workload.

\section*{\uppercase{Acknowledgements}}
This work was supported by the funding of the German Federal Ministry of Education and Research (BMBF) in the scope of the Software Campus program. Google Cloud credits in this work were provided by the \textit{Google Cloud Research Credits} program with the award number NH93G06K20KDXH9U.

\bibliographystyle{apalike}
{\small
\bibliography{bib}}

\begin{thebibliography}{}

\bibitem[Alam et~al., 2015]{analysis2011}
Alam, M., Shakil, K.~A., and Sethi, S. (2015).
\newblock Analysis and clustering of workload in google cluster trace based on
  resource usage.

\bibitem[Carreira et~al., 2019]{carreira2019cirrus}
Carreira, J., Fonseca, P., Tumanov, A., Zhang, A., and Katz, R. (2019).
\newblock Cirrus: A serverless framework for end-to-end ml workflows.
\newblock In {\em Proceedings of the ACM Symposium on Cloud Computing}, pages
  13--24.

\bibitem[Chadha et~al., 2021]{arch-specific}
Chadha, M., Jindal, A., and Gerndt, M. (2021).
\newblock Architecture-specific performance optimization of compute-intensive
  faas functions.
\newblock In {\em 2021 IEEE 14th International Conference on Cloud Computing
  (CLOUD)}, pages 478--483.

\bibitem[Cortez et~al., 2017]{miccrosoft_traces}
Cortez, E., Bonde, A., Muzio, A., Russinovich, M., Fontoura, M., and Bianchini,
  R. (2017).
\newblock Resource central: Understanding and predicting workloads for improved
  resource management in large cloud platforms.
\newblock In {\em Proceedings of the 26th Symposium on Operating Systems
  Principles}, SOSP '17, page 153–167, New York, NY, USA. Association for
  Computing Machinery.

\bibitem[Elgamal et~al., 2018]{costless}
Elgamal, T., Sandur, A., Nahrstedt, K., and Agha, G. (2018).
\newblock Costless: Optimizing cost of serverless computing through function
  fusion and placement.
\newblock {\em CoRR}, abs/1811.09721.

\bibitem[Espe. et~al., 2020]{espe_closer20}
Espe., L., Jindal., A., Podolskiy., V., and Gerndt., M. (2020).
\newblock Performance evaluation of container runtimes.
\newblock In {\em Proceedings of the 10th International Conference on Cloud
  Computing and Services Science - CLOSER,}, pages 273--281. INSTICC,
  SciTePress.

\bibitem[Fan. et~al., 2020]{closer20}
Fan., C., Jindal., A., and Gerndt., M. (2020).
\newblock Microservices vs serverless: A performance comparison on a
  cloud-native web application.
\newblock In {\em Proceedings of the 10th International Conference on Cloud
  Computing and Services Science - CLOSER,}, pages 204--215. INSTICC,
  SciTePress.

\bibitem[{Gao} et~al., 2020]{9209730}
{Gao}, J., {Wang}, H., and {Shen}, H. (2020).
\newblock Machine learning based workload prediction in cloud computing.
\newblock In {\em 2020 29th International Conference on Computer Communications
  and Networks (ICCCN)}, pages 1--9.

\bibitem[GoogleCloud, 2016a]{bigquery}
GoogleCloud (2016a).
\newblock Bigquery documentation.
\newblock Technical report.

\bibitem[GoogleCloud, 2016b]{dataproc:google}
GoogleCloud (2016b).
\newblock What is dataproc?
\newblock Google cloud documentation.
\newblock Posted at
  \url{https://cloud.google.com/dataproc/docs/concepts/overview}.

\bibitem[Guo et~al., 2019]{alibaba_traces}
Guo, J., Chang, Z., Wang, S., Ding, H., Feng, Y., Mao, L., and Bao, Y. (2019).
\newblock Who limits the resource efficiency of my datacenter: An analysis of
  alibaba datacenter traces.
\newblock In {\em Proceedings of the International Symposium on Quality of
  Service}, IWQoS '19, New York, NY, USA. Association for Computing Machinery.

\bibitem[Hellerstein, 2010]{clusterdata:Hellersetein2010}
Hellerstein, J.~L. (2010).
\newblock {Google} cluster data.
\newblock Google research blog.
\newblock Posted at
  \url{http://googleresearch.blogspot.com/2010/01/google-cluster-data.html}.

\bibitem[Kunde and Mukherjee, 2015]{shaju1}
Kunde, S. and Mukherjee, T. (2015).
\newblock Workload characterization model for optimal resource allocation in
  cloud middleware.
\newblock In {\em 4th SAC 15: Proceedings of the 30th Annual ACM Symposium on
  Applied Computing}, page 442–447.

\bibitem[Minet et~al., 2018]{8450304}
Minet, P., Renault, Ã., Khoufi, I., and Boumerdassi, S. (2018).
\newblock Analyzing traces from a google data center.
\newblock In {\em 2018 14th International Wireless Communications Mobile
  Computing Conference (IWCMC)}, pages 1167--1172.

\bibitem[Pacheco-Sanchez et~al., 2011]{shaju5}
Pacheco-Sanchez, S., Casale, G., Scotney, B., McClean, S., Parr, G., and
  Dawson, S. (2011).
\newblock Markovian workload characterization for qos prediction in the cloud.
\newblock In {\em 2011 IEEE 4th International Conference on Cloud Computing},
  pages 147--154.

\bibitem[Perennou et~al., 2019]{shaju6}
Perennou, L., Callau-Zori, M., Lefebvre, S., and Chiky, R. (2019).
\newblock Workload characterization for a non-hyperscale public cloud platform.
\newblock In {\em 2019 IEEE 12th International Conference on Cloud Computing
  (CLOUD)}, pages 409--413.

\bibitem[{Rasheduzzaman} et~al., 2014]{6779441}
{Rasheduzzaman}, M., {Islam}, M.~A., {Islam}, T., {Hossain}, T., and {Rahman},
  R.~M. (2014).
\newblock Task shape classification and workload characterization of google
  cluster trace.
\newblock In {\em 2014 IEEE International Advance Computing Conference (IACC)},
  pages 893--898.

\bibitem[Reiss et~al., 2012]{10.1145/2391229.2391236}
Reiss, C., Tumanov, A., Ganger, G.~R., Katz, R.~H., and Kozuch, M.~A. (2012).
\newblock Heterogeneity and dynamicity of clouds at scale: Google trace
  analysis.
\newblock In {\em Proceedings of the Third ACM Symposium on Cloud Computing},
  SoCC '12, New York, NY, USA. Association for Computing Machinery, Association
  for Computing Machinery.

\bibitem[Wilkes, 2011]{clusterdata:Wilkes2011}
Wilkes, J. (2011).
\newblock More {Google} cluster data.
\newblock Google research blog.
\newblock Posted at
  \url{http://googleresearch.blogspot.com/2011/11/more-google-cluster-data.html}.

\bibitem[Wilkes, 2020a]{clusterdata:Wilkes2020a}
Wilkes, J. (2020a).
\newblock {Google} cluster-usage traces v3.
\newblock Technical report, Google Inc., Mountain View, CA, USA.
\newblock Posted at
  \url{https://github.com/google/cluster-data/blob/master/ClusterData2019.md}.

\bibitem[Wilkes, 2020b]{clusterdata:Wilkes2020}
Wilkes, J. (2020b).
\newblock Yet more {Google} compute cluster trace data.
\newblock Google research blog.
\newblock Posted at
  \url{https://ai.googleblog.com/2020/04/yet-more-google-compute-cluster-trace.html}.

\end{thebibliography}

\end{document}